\documentclass[pdflatex,sn-nature]{sn-jnl}

\usepackage{graphicx}%
\usepackage{multirow}%
\usepackage{amsmath,amssymb,amsfonts}%
\usepackage{amsthm}%
\usepackage{mathrsfs}%
\usepackage[title]{appendix}%
\usepackage{xcolor}%
\usepackage{textcomp}%
\usepackage{manyfoot}%
\usepackage{booktabs}%
\usepackage{listings}%

\usepackage{color,soul}
\definecolor{lightgray}{rgb}{0.83, 0.83, 0.83}

\usepackage{lmodern}
\usepackage{anyfontsize}
\usepackage{amsfonts}
\usepackage{url}
\usepackage{booktabs}
\usepackage{amsmath}
\usepackage{xspace}
\usepackage{multirow}
\usepackage{amssymb}
\usepackage{wrapfig}
\usepackage[]{booktabs}
\usepackage{algorithm}
\usepackage[noend]{algpseudocode}
\renewcommand{\algorithmicrequire}{\textbf{Input:~}}
\renewcommand{\algorithmicensure}{\textbf{Output:~}}

\newcommand\ggx[1]{#1}

\usepackage{array}
\usepackage{color}
\usepackage{listings}
\usepackage{placeins}
\usepackage{enumitem}

\algnewcommand{\algorithmicforeach}{\textbf{for each}}
\algdef{SE}[FOR]{ForEach}{EndForEach}[1]
  {\algorithmicforeach\ #1\ \algorithmicdo}
  {\algorithmicend\ \algorithmicforeach}

\renewcommand{\algorithmicrequire}{\textbf{Input:~}}
\renewcommand{\algorithmicensure}{\textbf{Output:~}}

\newcommand\acro{{\sc{DiFuseR\xspace}\xspace}\xspace}
\newcommand\acros{{\sc{HyperFuseR\xspace}\xspace}\xspace}

\newcommand\acrod{{\sc{DiFuseR\xspace}\xspace}\xspace}

\usepackage[caption=false,font=normalsize,labelfont=sf,textfont=sf]{subfig}

\begin{document}

\title[{\sc DiFuseR}]{{\sc DiFuseR}: A Distributed Sketch-based\\ Influence Maximization Algorithm for GPUs}

\author*[1,2]{\fnm{G\"okhan} \sur{G\"okt\"urk}}\email{ggokturk@sabanciuniv.edu}

\author[1,2]{\fnm{Kamer} \sur{Kaya}}\email{kaya@sabanciuniv.edu}

\affil[1]{\orgdiv{Faculty of Engineering and Natural Sciences}, \orgname{Sabancı University}, \orgaddress{\city{Istanbul}, \postcode{34956}, \country{Turkey}}}

\affil[2]{\orgdiv{Center of Excellence in Data Analytics}, \orgname{Sabancı University}}

\abstract{Influence Maximization~(IM) aims to find a given number of ``seed" vertices that can effectively maximize the expected spread under a given diffusion model. 
Due to the NP-Hardness of finding an optimal seed set, approximation algorithms are often used for IM. However, these algorithms require a large number of simulations to find good seed sets. 
In this work, we propose \acrod, a blazing-fast, high-quality IM algorithm that can run on multiple GPUs in a distributed setting. \acrod is designed to increase GPU utilization, reduce inter-node communication, and minimize overlapping data/computation among the nodes. Based on the experiments with various graphs, containing some of the largest networks available, and diffusion settings, the proposed approach is found to be $3.2\times$ and $12\times$ faster on average on a single GPU and 8 GPUs, respectively. It can achieve up to $8\times$ and $233.7\times$ speedup on the same hardware settings. Furthermore, thanks to its smart load-balancing mechanism, on 8 GPUs, it is on average $5.6\times$ faster compared to its single-GPU performance. 
}
\keywords{Influence Maximization, Graph Processing, Count-Distinct Sketch, Error-Adaptive Cardinality Estimation.}



\maketitle
\newpage
\section{Introduction}

Influence Maximization~(IM), introduced by Kempe et al.~\cite{kempe2003maximizing}, is the problem of identifying a set of individuals in a social network who, when influenced/activated, can effectively propagate a particular message or influence to a larger portion of the network. IM aims to find a small group of ``seed" vertices that can effectively maximize the expected spread of a message or influence throughout the network under a given diffusion model.

IM is extremely useful in the context of marketing, advertising, pandemic analysis, and political campaigns. It is a powerful tool within the context of social media, where the rapid spread of information and ideas can have a profound impact on public discourse and decision-making. In addition to its practical applications, IM is also of theoretical interest to researchers studying social networks and the spread of information.
However, identifying a good set of seed nodes to efficiently initiate and perform the spread is a  challenging task, especially for large networks. The quality of the seed set is particularly important when the spread of information is a time-critical task, which makes it hard to get high-quality sets, or when using large seed sets is not feasible due to cost constraints.

Due to the NP-Hardness of finding an optimal seed set\cite{kempe2003maximizing}, approximation algorithms are often used for IM. The randomized greedy approach is one of the most frequently applied algorithms for IM. The time complexity of the randomized algorithm, with an influence score estimate $\sigma$, running $\mathcal{R}$ simulations, and selecting $K$ seed vertices is $\mathcal{O}(K\mathcal{R}n\sigma)$ for a graph with $n$ vertices. Although these greedy Monte-Carlo algorithms perform well in terms of seed set quality, they can be impractical for real-life networks containing millions of vertices and edges due to expensive simulation costs. 
Due to this reason, many heuristics and proxy methods have been proposed in the literature.\cite{MixGreedy, narayanam2010shapley, kimura2007extracting, chen2010PMIA,chen2010LDAG, kim2013scalable, cohen2014sketch, goyal2011simpath, jung2012irie,cheng2014imrank,liu2014influence,galhotra2016holistic}. In this work;

\begin{itemize} 
  \item We propose \acrod, a blazing-fast, high-quality IM algorithm that can leverage multiple nodes and GPUs in a distributed setting. \acrod is designed to be simple, to increase GPU utilization, reduce inter-node and inter-GPU communication, and minimize overlapping data/computation between nodes. It uses {\tt MPI} to distribute the work to multiple nodes and its inner kernels are written in {\tt CUDA} to leverage GPUs. 

  \item \acro leverages data sketches to work with estimations instead of computing exact values as suggested by~\cite{hyperfuser} for CPUs. We design, implement, and utilize a GPU-specialized variant of the count-distinct sketches. The sketch values are used as the estimates of the number of vertices that will be  activated in the simulations when a vertex is inserted in the seed set.  Thanks to the suitability of these sketches on GPUs, this not only reduces the amount of computation but also keeps the device utilization high.\looseness=-1  
  
  \item \acro samples the edges as they are being traversed within simulations. Therefore within a simulation, the sampling and diffusion processes are fused for a single-edge traversal; a technique borrowed from the literature.\looseness=-1 

  \item We propose the {\em fusing-aware sample-space tasking} (FASST) scheme to partition the graph's edges among the devices. With FASST, \acro can exploit the hash-based pseudo-random number generation used for the Monte-Carlo process and minimize the data overlap among the devices. More formally, for an edge $e$, FASST aims to maximize the number of Monte-Carlo simulations assigned to the same GPU warp and containing $e$.
  
  \item On a distributed setting, \acrod can be $167\times$ than its nearest competitor while providing $12\times$ speedup on average. Furthermore, it achieves up to $20.74\times$ ($7\times$ on average) speedup with eight GPUs compared to a single GPU. Furthermore, it is able to process some of the largest networks available.

  \end{itemize}
  
  The paper is organized as follows: The notation used in the paper is given in Section~\ref{sec:back}. A background on IM is also provided in this section. Section~\ref{sec:acro} describes the building blocks of \acro. The main backbone of \acrod designed to employ multiple machines and GPUs and fusing-aware sample-space tasking scheme are described in Section~\ref{sec:difuser}. The experimental results are given in Section~\ref{sec:exps}. Section~\ref{sec:rel} summarizes the related work, and Section~\ref{sec:conc} concludes the paper.
  
  \section{Notation and Background}\label{sec:back}
  
  Let $G = (V,E)$ be a network graph where the $n$ vertices in $V$ represent the agents and $m$ edges in $E$ represent the relations among them. For {\em directed} graphs, an edge $(u,v) \in E$ is an {\em incoming} edge for $v$ and an {\em outgoing} edge of $u$. The {\em incoming} neighborhood of a vertex $v \in V$ is denoted as $\Gamma^-_{G}(v) = \{u: (u,v) \in E\}$. Similarly, the {\em outgoing} neighborhood of a vertex $v \in V$ is denoted as $\Gamma^+_{G}(v) = \{u: (v,u) \in E\}$. 
  The reachability set of a vertex $v \in V$ is the union of all its \ggx{(both directly and indirectly)} connected vertices (starting with the outgoing edges of $v$). 
  
  A graph $G' = (V',E')$ is a sub-graph of $G$ if $V' \subseteq V$ and $E' \subseteq E$. The diffusion probability on the edge $(u, v) \in G$ is noted as $w_{u,v}$, where $w_{u,v}$ can be determined either by the diffusion model or \ggx{can be given according to the relationship between $u$ and $v$.}
  In practice, $w_{u,v}$ can be determined by the strength of $u$ and $v$'s relationship~{\cite{kempe2003maximizing}}. The notation we use in the paper is summarized in Table~\ref{tab:notation}.
  
  \begin{table}[!ht]

        \centering
        \caption{Notation used in the paper}
        \label{tab:notation}
        \begin{tabular}{|l|p{0.68\linewidth}|}
            \hline
            Variable & Definition  \\
            \hline
            $G = (V,E)$     & Graph $G$ with vertices $V$ and edges $E$ \\
            $\Gamma^+_{G}(v)$ & {\em outgoing} neighborhood of a vertex (directed graphs) \\
            $\Gamma^-_{G}(v)$ & {\em incoming} neighborhood of a vertex (directed graphs) \\
            $w_{u,v}$       & Probability of $u$ directly influencing $v$ \\
            \hline\hline
            $S$             & Seed set to maximize influence\\
            $K$             & Size of the seed set\\
            $\mathcal{R}$   & Number of Monte-Carlo simulations performed\\
            $\sigma_{G}(S)$& Influence score of $S$ in $G$, i.e., expected number of vertices reached from $S$ in $G$\\
            $\sigma_{G}{(S,v)}$          & Marginal influence gain by adding vertex $v$ to seed set $S$\\
    \hline\hline
            $h(u,v)$        & Hash function for edge $(u,v)$\\
            $h_j(x)$        & $j$'th hash function for number $x$\\
            $h_{max}$       & Maximum value hash function $h$ can return\\
            $X_r$           & Random number/hash generated for simulation $r$  \\
            $\oplus$ & XOR operator\\
            \hline\hline

          ${\cal J}$ & Number of registers for sketches\\
          $clz(.)$ & Count leading zeros function\\
          $H(.)$ & Harmonic mean function\\
            $\mu$ & Number of devices/partitions\\
            \hline
        \end{tabular}
    \end{table}

    \subsection{Influence Maximization Problem}

    IM problem aims to find a seed set $S \subseteq V$ among all possible size $K$ subsets of $V$ that maximizes an {\em influence spread function} $\sigma_{G,M}$  on $G$ under a diffusion model $M$ when the diffusion process is initiated from $S$. 
    The influence spread function $\sigma_{G,M}(\cdot)$ computes the {\em expected} number of agents~(or nodes, vertices, etc.) influenced~(activated) through a diffusion model $M$. For the sake of simplicity, we drop $M$ from the notation; in the rest of the text, $\sigma_{G}$ refers to $\sigma_{G,M}$. Some of the popular diffusion models for IM in the literature are {\em independent} and {\em weighted cascade}~(IC and WC),  and {\em linear threshold}~(LT)~\cite{kempe2003maximizing}. 
    
    \begin{figure}[!ht] 
        \centering
      \subfloat[\small{IC}\label{fig:ic}]{%
           \includegraphics[width=0.25\linewidth]{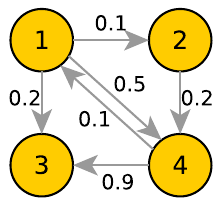}}\hspace*{4ex}
      \subfloat[\small{WC}\label{fig:wc}]{%
            \includegraphics[width=0.25\linewidth]{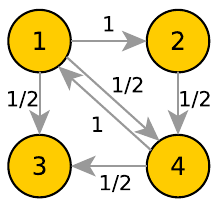}}
        \\
      \caption{{\protect\subref{fig:ic} 
    The directed graph $G = (V, E)$ for IC with independent diffusion probabilities. 
    \protect\subref{fig:wc}
    The directed graph for WC is obtained by setting the diffusion probabilities of incoming edges to $1 / |\Gamma^-_G(v)|$ for each vertex $v \in V$. 
      }}
    \end{figure}

    \begin{itemize}
      \setlength\itemsep{1em}

    \item The {\bf Independent Cascade} model, which we focus on in this work, runs in rounds and activates a vertex $v$ in the current round if one of $v$'s incoming edges $(u, v)$ is used during the diffusion round, which happens with the activation probability $w_{u, v}$, given that $u$ has already been influenced in the previous rounds. The activation probabilities are independent (from each other and previous activations) in the {\em independent cascade} model. A toy graph with activation probabilities on the edges is shown in Fig.~\ref{fig:ic}.
    In theory, there can exist parallel and independent $(u, v)$ edges in $E$. In practice, they are merged to a single $(u,v)$ edge with compound probability via preprocessing. For example, the probability of vertex {\em u} activating vertex {\em v} through two parallel edges with $w=0.1$ is $w_{u,w} = 1-(1- 0.1)^2 = 0.19$.
    
    \item The {\bf Weighted Cascade}  model is a variant of IC and uses the structural properties of vertices to set the edge weights as shown in Fig.~\ref{fig:wc}.
    The method, as described by Kempe~et~al.\cite{kempe2003maximizing}, sets $w_{u, v} = 1 / d_v$ where $d_v$ is the number 
    of incoming edges of $v$~(which in the original graph is equal to $\Gamma^-_G(v)$).
    Therefore, if $v$ has $\ell$ neighbors activated in the last round, its probability of activation in the new round is $1-( 1-1 / d_v)^\ell$. 
    
    \item{\bf Linear threshold} generalizes the IC model and activates the vertex $v$ once the cumulative activation coming from its neighbors exceeds a given threshold $\theta_v$. 
    All the $(u, v)$ edges with active $u$ vertices are taken into account in the process. Vertex $v$ is activated when the total activation probability through these edges exceeds $\theta_v$~\cite{kempe2003maximizing}.  

    \end{itemize}
    
    In this paper, we focus on the Independent Cascade model, but similar approaches can be applied to other models as well. The IM problem is NP-hard under the cascade and linear threshold models. This being said, the influence function is { {non-negative,}} monotone and sub-modular, which means that adding a single vertex to the seed set at hand can only increase the overall influence and decreases the marginal influence scores for the remaining vertices that are not in the seed set. Due to these properties, the influence score of a greedy solution, which always adds the most promising vertex with the highest marginal gain to a seed set of final size $K$ is at least $1-(1-1/K)^K \geq 63\%$ of the optimal solution~\cite{nemhauser1978analysis}. 
    
    \subsection{Hash-based Fused Sampling}

    There are various problems with randomized algorithms leveraging graph sampling; the main one is the extra overhead due to randomization. Moreover, the cost of the traditional pipeline, i.e., sample, organize, store, and process, drastically reduces the performance. To overcome these problems, G\"{o}kt\"{u}rk and Kaya proposed the hash-based fused-sampling approach. Additionally, they proposed faster single-node, CPU-based IM approximation algorithms~\cite{infuser}. They remodeled the existing, traditional algorithms by deciding the samples on-the-fly, without explicitly computing the samples in a pre-processing step. Note that unlike \acro, these algorithms run on CPUs, construct the reachability sets, and use exact cardinality values instead of estimating them.
    
\begin{figure}[!ht] 
    \captionsetup[subfigure]{justification=centering}
        \centering
        \subfloat[\label{fig:sims}]{%
            \includegraphics[width=0.30\linewidth]{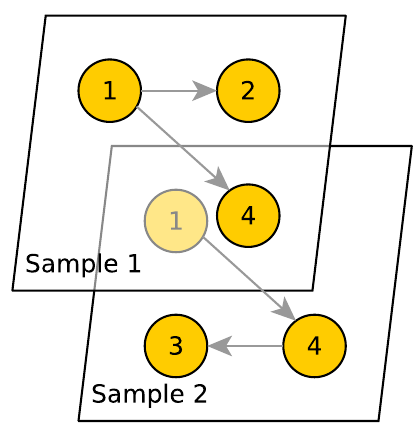}
        } 
        \subfloat[\label{fig:fused}]{%
         \includegraphics[width=0.30\linewidth]{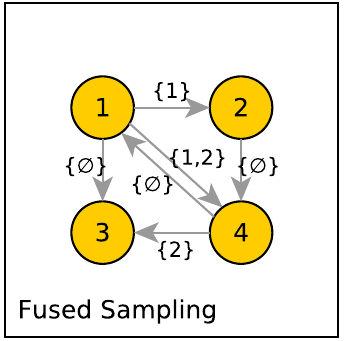}}
      \caption{\small{
      \protect\subref{fig:sims} Two sampled subgraphs of a toy graph with 4 vertices and 6 edges.
      \protect\subref{fig:fused} The simulations performed are fused with sampling. Each edge is labeled with the corresponding sample/simulation IDs. 
      }}
      \label{fig:traversal} 
    \end{figure}

    Under the cascade models, the randomized greedy approximation requires sampling subgraphs from $G = (V, E)$ to simulate the diffusion process. As mentioned above, sampling is expensive and its cost can dominate the runtime. Fusing eliminates the necessity of the creation and storage of the sample subgraphs. That is when an edge of $G$ is being processed, it is processed for all possible samples as in Fig.~\ref{fig:traversal}. Then, it is decided to be {\it sampled} or {\it skipped} depending on the outcome of the random value for each sample/simulation. 
    This allows the algorithm to access each edge only a few times instead of once for every simulation, which reduces the pressure on the memory subsystem and increases the computation/communication ratio. 
    
    Hash-based sampling generates the random values with good hash functions. Given a graph $G = (V, E)$, for an edge $(u, v) \in E$, the hash function used in~\cite{infuser} is:
    \begin{equation}
        \label{eq:hash}
        h(u,v) = \mbox{{\sc Murmur3}}(u||v)~mod~2^{31}  
    \end{equation}
    where $||$ is the concatenation operator. Although this function generates a unique hash value for each edge, different simulations require different probabilities. To have them, a set of random values, ${\cal X}$, is generated such that $${\cal X} = \{X_1, X_2,\ldots, X_{\cal R}\}$$ where each $X_r \in [0, h_{max}]$, associated with simulation $r$, is uniformly randomly chosen from the interval~\cite{infuser}. Then the sampling probability of $(u, v)$ for simulation $r$, $P(u, v)_r$, is computed as 
    \begin{equation}
        \label{eq:hash_prob}
        P(u,v)_r = \frac{X_r \oplus h(u,v)}{h_{max}}.
    \end{equation} where $\oplus$ is the XOR operator.

    The edge $(u,v)$ exists in the sample $r$ if and only if  ${P}(u,v)_r$ is smaller than the edge threshold $w_{u,v}$. One of the benefits of this approach is that an edge can be sampled using a single XOR and compare-greater-than operation. Moreover, the corresponding control flow branch overhead can be removed using {\em select} (predicate) instructions by the CUDA compiler.
    
\subsection{Flojet-Martin Sketches}
Flajolet-Martin~(FM) sketch~\cite{flajolet1985probabilistic} is a probabilistic data structure used for estimating the number of distinct elements in a data stream. The estimation process leverages the idea of using {\em rarity} which is obtained by counting the maximum number of leading zeros on the hash values of stream elements and using the maximum rarity to estimate the stream cardinality using only a little memory. Formally, the sketch maintains a set of registers, each corresponding to a hash function. When a new element is processed, it updates a register if the number of leading zeros in the element's corresponding hash value is bigger. The estimated number of distinct elements is then calculated by taking the mean of the registers.

FM sketches have several useful properties, including a low memory footprint, good enough accuracy for a wide range of applications, and the ability to perform approximate set union and intersection operations, which we exploit for IM. Using a sketch reduces the time and space complexity of these set operations to a constant number of operations, which is particularly useful in Monte-Carlo-based IM, since the algorithm is already randomized and non-optimal, an exact number is not required, and a good estimate may be sufficient. Here is how \acrod leverages FM sketches:

\begin{itemize}
  \setlength\itemsep{0.5em}
\item{{\bf Initialization:}}
For every vertex in $G = (V, E)$ and for every simulation pair, \acrod employs an FM sketch. A sketch initializes all its registers to zero in the beginning. Since each vertex $u \in V$ is a member of its reachability set~(with distance $0$), we immediately add $u \in V$ to its corresponding sketches. The initialization for $u$'s registers can be formulated as follows:
\begin{equation}
    \label{eq:sketch-init}
    M_u[j] = \text{\sc clz}(h_j(u)),~1 \leq j \leq {\cal J}
\end{equation} where $h_j$ is the $j$th hash function, $clz(y)$ returns the number of leading zeros in $y$ which is a member of the range of $h_j$, and ${\cal J}$ is the number of sketch registers per vertex.

\item{{\bf Adding Items:}}

For a sketch with ${\mathcal J}$ registers, the impact of adding an item $v \in V$ is shown in~\eqref{eq:sketch-add}:
\begin{equation}
    \label{eq:sketch-add}
    M[j] = \text{\sc max}(M[j], \text{\sc clz}(h_j(v)),~1 \leq j \leq {\cal J}
\end{equation}

\item{{\bf Union:}}
Two given FM sketches $M_u$ and $M_v$ can be merged, i.e., their union $M_{uv}$ can be computed by taking the pairwise maximums of their registers. Formally; 
\begin{equation}
\label{eq:sketch-merge}
    M_{uv}[j] = \text{\sc max}(M_u[j], M_v[j]),~1 \leq j \leq {\cal J}.
\end{equation} 
In traditional IM algorithms, while simulating a diffusion process, one needs to combine the reachability sets of vertices $u$ and $v$ if the edge $(u,v)$~(or $(v, u)$) becomes active, i.e., if the edge is sampled for this simulation. To choose a seed vertex, this union operation may need to be performed for each sampled edge in each simulation. Hence, its overhead incurs an important bottleneck for IM computations. However, in the end, it is not the set that is used but only its cardinality. In \acrod, we exploit this fact and perform the sketch merge given in~\eqref{eq:sketch-merge} by only using fewer, $\mathcal{O}(J)$, of operations. 

\item{\bf{Cardinality Estimation:}}
With a single register, the cardinality estimation can be done by computing the power $2^r$ where $r$ is the value in the register and applying a correction factor. With multiple registers, the average of the register values can be used to estimate the cardinality, and the result is divided by a correction factor $\phi$ to fix the error due to hash collisions. Flajolet et al. \cite{flajolet1985probabilistic} suggest $\phi \approx 0.77351$ for $\mathcal{J}>=16$ registers. The estimated cardinality $e$ is computed as
\begin{equation}
    \label{eq:sketch-estimate}
    e = 2^{\bar{M}}/\phi
\end{equation} 
where $\bar{M} = {\text{\sc avg}}_j\{M[j]\}$ is the mean of the register values.
\end{itemize}


\section{Sketch-based Influence Maximization on GPU}\label{sec:acro}

The traditional approach for Monte-Carlo-based IM starts with an empty seed set and finds the {\em exact} marginal gains for each vertex $v  \in V$, which is the expected number of vertices activated when $v$ is inserted into the seed set. Then in the next iteration, the second seed vertex $u \neq v$ is found by computing the expected number of {\em new} activations. This process continues until the desired number of seed vertices is obtained. For an {\em exact} computation of marginal gains at each iteration, one needs to keep the state (i.e., activated or not) of each vertex for each simulation. This information can be considered as the reachability set of the seed set for each sampled graph. However, maintaining these sets is infeasible for large graphs. With $\mathcal{R}$ Monte-Carlo simulations and one-hot encoded {\em membership} vectors with constant insertion time for efficiency, the space complexity becomes $\mathcal{O}(n^2\mathcal{R})$, and the time complexity of a merge operation becomes $\mathcal{O}(n)$ where $n$ is the number of vertices. 
\(\mathcal{O}(n^2\mathcal{R})\) arises because each of the \(\mathcal{R}\) Monte Carlo simulations requires storing \(n\) one-hot encoded vectors, each of length \(n\), leading to a total space requirement of \(\mathcal{O}(n^2\mathcal{R})\). \(\mathcal{O}(n)\) for the merge operation comes from the need to process \(n\) vertices individually in the merge step, with each vertex's processing taking constant time.
Storing only reachable elements can require total influence at most; it is possible to store $\mathcal{O}(\sigma)$ elements instead of $n$ elements for each vertex in all simulations.   
Therefore, the space complexity can be reduced to $\mathcal{O}(n{\sigma}\mathcal{R})$ by using sparse vectors~(i.e., sets) or specialized data structures such as disjoint sets. However, it will still be infeasible for large-scale graphs, considering that our target architecture is memory-restricted GPUs. 

In this work, to solve these problems, (1) we propose a modified sketch data structure specially designed for GPUs, and (2) this changes the structure of an efficient IC implementation for Influence Maximization. For instance, in this work, the Greedy algorithm is heavily optimized for GPUs. (3) To be more efficient, novel sampling-aware data distribution, task assignment, and scheduling methods have been designed and implemented. All these contributions make this work superior to the literature and much faster while being memory efficient.

\subsection{FM Sketches on GPU}

The count-distinct sketch used in \acros~\cite{hyperfuser} is designed for efficiency and optimized for AVX2/AVX512 capable processors. Due to the inherent architectural differences between AVX acceleration and CUDA-capable devices, in this work, we needed to explore a new sketch for GPUs.

As mentioned before, each sketch uses a set of registers for bookkeeping. An 8-bit register stores only a single number of leading zeros, i.e., one $clz$ output. However, this will be a waste of space considering that 7 bits are enough to store the $clz$ for 64-bit hash functions. The additional space can indeed be used to increase the seed set quality and improve the execution time: we encoded the {\em visited} information on the vertices inside the remaining bits of these sketch registers. This allows finding the sample/simulation IDs for each vertex $u$ in which $u$ is active without using extra memory. Note that when a vertex is visited in a current iteration of the simulation, it is not required to check it in the next one. 

This and utilizing the extra bit allow us to significantly reduce the computation and memory accesses while rebuilding sketches by performing early exits within a simulation and to utilize conditional move statements while comparing registers.

Since the simulations can be performed independently from each other, the multi GPU-setting does not change the way \acrod processes them compared to the single GPU setting. However, the notation slightly changes. For instance, to extract values from the sketches, we use
\begin{equation} \label{eq:gpu-sketch-int}
    M_u[j] = \text{\sc H}(\text{\sc clz}(h_{\tau J+j})),~1 \leq j \leq {\cal J}
\end{equation} 
for $0 \leq j < {\cal J}$,  $H$ is the harmonic mean function, $clz$ is count leading zeros function, and $h_j$ is the to $j$th hash function. Here $\tau$ is the device ID which will be explained later while describing the proposed distributed multi-GPU solution. For a single-GPU execution, $\tau$ is 0. Similarly, we use ${\cal J}$ as the number of simulations/registers {\em per device}. For a single-GPU execution, it is equal to the total number of simulations. Using the harmonic mean for aggregation helps us to remove the extreme values before they affect the rest without any post-processing. A similar approach has been applied in {\sc HyperLogLog++}~\cite{hllpp} which removes $30\%$ of the largest values to improve the sketch accuracy. 

For clarity, we kept the pseudocodes in the paper as close to the implementation as possible. For instance, {\tt blockIdx} and {\tt threadIdx} are the built-in CUDA variables that store the coordinates of the current block and current thread, respectively. Similarly, the built-in variable {\tt gridDim} specifies the grid dimensions. The algorithm for initializing sketches in the multi-GPU setting is given in Algorithm~\ref{algo:fill-sketches}. The memory layout of the registers is organized in a way that all the registers of a vertex $u$ are adjacent to each other in the order of increasing simulation IDs. Hence, to achieve a better memory access pattern and a better load balance, most of the \acrod kernels assign a vertex $u \in V$ to a block $blk$ and all the register operations for $u$ to the threads of $blk$. 

One important detail that needs to be explained is that line~\ref{ln:exit} of the Algorithm~\ref{algo:fill-sketches} exploits the extra bit and performs an early exit as described before. That is if $u$ is visited for a simulation, the corresponding registers, which are set to $-1$ at the time $u$ is visited, will not be used again (also for the next seed set selection). This idea will be reused also in other kernels.   

\renewcommand{\baselinestretch}{0.95}
\begin{algorithm}[!ht]
\small
\caption{\sc{Fill-Sketches}($G,M,{\cal J}, \tau$)}
\label{algo:fill-sketches}
\algorithmicrequire{$G = (V,E)$: the influence graph
\\\hspace*{6.6ex}$M$: sketch vectors of vertices
\\\hspace*{6.7ex}${{\cal J}}$: number of registers per vertex
\\\hspace*{6.7ex}$\tau$: device id
\\\hspace*{6.7ex}$\mu:$ number of devices
}
\\\algorithmicensure{$M$: updated sketch vectors
}
\begin{algorithmic}[1]
    \State {$u \leftarrow blockIdx.x$} \Comment{Vertex assigned to the thread}
    \State {$j \leftarrow threadIdx.x + \tau\mathcal{R}/\mu $} \Comment{Sample/simulation ID offset}
    \State {$stride  \leftarrow gridDim.x$}    
    \While {$u < |V|$}
        \If{$M_u[j] \neq -1 $}\label{ln:exit}
            \State{\bf{continue}}
        \EndIf

        \State $M_u[j] \leftarrow clz(h_j(u))$

        \State {$u \leftarrow u + stride$}
    \EndWhile
\end{algorithmic}
\end{algorithm}
\renewcommand{\baselinestretch}{1}

\subsection{Updating Sketches on GPU}

\acrod performs multiple simulations efficiently on the GPU at once via consecutive single-step iterations. Each iteration extends the diffusion path in the network and relays one level of cardinality information. That is each vertex $u$, gathers the estimated cardinality of the vertices in its outgoing neighborhood for each sample, and accumulates them in its corresponding registers via sketch merge operations. The pseudocode of a single iteration is given in Algorithm~\ref{algo:simulate-sketches}. The full execution of this process requires at most $d$ iterations where $d$ is the maximum diameter among the sampled graphs. Considering that the diameters of these graphs can vary significantly, \acrod seems to perform many unnecessary iterations. However, since multiple updates for different simulations are condensed within a warp and we have early exits in action, the number of redundant operations is much less and the benefits of this approach outweigh the overhead. 

\ggx{We note that \acrod uses a pull-based approach to diffuse cardinalities between vertices whereas other methods we compare ourselves to usually are push-based approaches\cite{gim,curipples,CELF,kempe2003maximizing}. This allows us to perform idempotent updates, avoiding atomic operations altogether. If we were to employ a pull-based sketch update step, atomic writes would have been required on every register of every visited neighbor. On the other hand, our pull-based approach can accumulate updates of its neighbors and update registers only once per depth iteration. Even if neighboring registers have been recently updated in the same iteration, we would be only pulling the next iteration's cardinality sketches.
}

\renewcommand{\baselinestretch}{0.95}
\begin{algorithm}[!ht]
\small
\caption{\sc{Simulate}($G,M,{\cal J}$)}
\label{algo:simulate-sketches}
\algorithmicrequire{$G = (V,E)$: the influence graph
\\\hspace*{6.6ex}$M$: sketch vectors of vertices
\\\hspace*{6.7ex}${{\cal J}}$: number of registers per vertex
}
\\\algorithmicensure{$M$: updated sketch vectors
}
\begin{algorithmic}[1]
    \State {$u \leftarrow blockIdx.x$}
    \State {$j \leftarrow threadIdx.x$}
    \State {$stride  \leftarrow gridDim.x$}
    \While {$u<|V|$}
        \If{$M_u[j] = -1 $} \State {\bf continue}
        \EndIf
        \For{$v \in \Gamma_G^+(u) $}
            \If{$P(u,v)_j < w_{u,v}$}                
                \State{$M_u[j] \leftarrow \text{\sc max}(M_u[j],M_v[j])$}
            \EndIf
        \EndFor
        \State {$u \leftarrow u + stride$}
    \EndWhile
\end{algorithmic}
\end{algorithm}
\renewcommand{\baselinestretch}{1}

\subsection{Influence Cascade on GPU}

Given the current seed set, almost all of the IM algorithms must find the set of influenced vertices for all samples. This allows the algorithm to first compute the current influence score and be ready to efficiently compute the marginal gain for the next seed candidate. This process, which basically requires multiple sampled-graph traversals, is called {\em Influence Cascade}. An independent traversal of each sampled graph using a separate queue for each worker is a trivial approach for performing the influence cascade on GPU. However, this approach requires $\mathcal{O}(|V|)$ storage per worker which makes it infeasible on memory-restricted GPUs. Furthermore, processing a single edge with a single thread during these traversals incurs path divergence within each warp. The divergent warp problem can be mitigated by processing a vertex per block/warp while processing the edges associated with that vertex with the warp threads. This technique is applied by {\sc gIM}, an IM tool working on a single GPU~\cite{gim}. However, it incurs limits to parallelism. In this work, we use  {\sc gIM} as a baseline. 

\renewcommand{\baselinestretch}{0.95}
\begin{algorithm}[!ht]
\small
\caption{\sc{Cascade}($G,M,{\cal J}, Q, \hat{Q}$)}
\label{algo:cascade-kernel}
\algorithmicrequire{$G = (V,E)$: the influence graph
\\\hspace*{6.6ex}$M$: sketch vectors of vertices
\\\hspace*{6.7ex}${{\cal J}}$: number of registers per vertex
\\\hspace*{6.7ex}$Q$: current queue
\\\hspace*{6.7ex}$\hat{Q}$: queue of the next iteration
}\\
\algorithmicensure{$M$: updated sketch vectors
}
\begin{algorithmic}[1]
    \State {$i \leftarrow blockIdx.x$}
    \State {$j \leftarrow threadIdx.x$}
    \State {$stride  \leftarrow gridDim.x$}
    \State {$lane  \leftarrow j~mod~warpSize$}
    \State {$rnd  \leftarrow rands[j]$}
    \While{$ i < size(Q) $}
        \State  $u \leftarrow Q[i]$
        \If {$M_v[j] \neq 1$} 
            \State {\bf continue} 
        \EndIf
        \For{$v \in \Gamma_G^+(u) $}
            \State{$flag \leftarrow 0$}
            \If{$P(u,v)_j < w_{u,v}$}                
            \State{$M_u[j] \leftarrow -1$}
                \State{$flag \leftarrow 1$}                    
            \EndIf
            \If {$lane = 0 \land \text{\sc any}(flag)=1$ }
                \State $\hat{Q} \leftarrow \hat{Q} \cup \{v\}$  \Comment{Atomic} \label{ln:atomic}
            \EndIf
        \EndFor
        \State $i \leftarrow i + stride$
    \EndWhile
    \State {\Return} $M$
\end{algorithmic}
\end{algorithm}
\renewcommand{\baselinestretch}{1}

In \acrod, we have used a different solution for avoiding divergence; the algorithm combines the (vertex) frontiers, i.e., takes their union, for each sample, and processes the vertices that appear in a frontier of at least one sample in parallel. This process is given in Algorithm~\ref{algo:cascade-kernel}. It is repeated until the convergence, the state with no update, is reached. This implies that all traversal levels have been processed for all samples. Similar approaches have been applied for different graph kernels such as centrality computation by Sarıyüce~et~al.~\cite{SariyuceSKC15}. 

Instead of using multiple queues, i.e., one queue per sample, \acrod uses a single, unified queue to store the distinct vertices to be traversed. For simplicity (at the implementation level), we have two unified queues; the first one, $Q$, stores the vertices that will be processed in the current kernel launch. The second unified queue, $\hat{Q}$, accumulates the vertices to be processed in the next iteration. At each iteration, we start a block for each vertex in the first segment~(with an upper limit on the number of concurrent blocks). Then each thread in the block processes the same edge for different samples. Finally, the vertices visited by the threads are accumulated and added to the latter segment of the queue to be processed in the next iteration. After the kernel finishes, we swap the first and latter segments and repeat until no more elements stay in the queue. 
\ggx{The queues in Algorithm~\ref{algo:cascade-kernel} contain only unique elements and are implemented as sets. Membership operations are performed on a 1-hot array for performance and consistency, while traversal operations are performed on a continuous array.} 

With unified queues, the only performance disrupting memory access is the atomic operations performed for adding a new vertex to the queue~(line~\ref{ln:atomic} of Algorithm~\ref{algo:cascade-kernel}). To reduce the overhead of atomic operations, we used a warp-synchronized per-thread variable $flag$, which reduces the number of insertions to $\hat{Q}$ to one per warp. In the algorithm, $any$ is the function that performs the warp-level reduction for $flag$. Although this approach can also be enhanced to block-level, such an extension increases the synchronization overhead.

Since the visited vertices (by the current seed set) in a sample cannot have an impact on the influence, as mentioned before, we set the sketch register of that sample to a hard-coded value, {\tt -1}, to label the vertex visited in the associated sample. 
With the proposed approach, we are able to process thousands of vertices in parallel without memory bottlenecks. In our preliminary experiments, the run-time analysis shows more than $80\%$ utilization of the streaming multiprocessors~(SMs) and the memory sub-system.

\section{\acrod: Distributed IM on GPUs}\label{sec:difuser}

\renewcommand{\baselinestretch}{0.95}
\begin{algorithm}[!ht]
\small
\caption{\sc{\acrod}(${G'}_{\tau} ,{\cal J}, K, \tau, \mu$)}
\label{algo:superfuser}
\algorithmicrequire{${G'}_{\tau} = (V,E_{\tau})$: device-local graph for $\tau$th device 
\\\hspace*{6.7ex}${{\cal J}}$: number of Sketch registers
\\\hspace*{6.7ex}$K$: Size of the seed set
\\\hspace*{6.7ex}$\tau$: Device/partition id
\\\hspace*{6.7ex}$\mu$: Number of devices
}
\\\algorithmicensure{$S$: Seed set
}
\begin{algorithmic}[1]
    \State {$S \leftarrow \varnothing$}\Comment{\parbox[t]{.37\linewidth}{Initialize the seed set as empty.}}
    \State {$score \leftarrow 0,~oldscore \leftarrow 0$}
    \State {$M \leftarrow ${\sc{zeros}} $ ( |V|, {\cal J})$}   
    \State {$M \leftarrow ${\sc{Fill-Sketches{\tt <<<}$B,{\cal J}$}{\tt >>>}}$(M, {\cal J}, \tau)$\Comment{\parbox[t]{.37\linewidth}{Initialize sketches by adding each vertex to its sketches.}}}\label{ln:1}
    \While {$M$ not converged}
    \State $M \leftarrow ${\sc Simulate {\tt<<<} $B,{\cal J}${\tt >>>} }$({G'}_{\tau},M,{\cal J})$\label{ln:2}\Comment{\parbox[t]{.37\linewidth}{Merge each vertex's sketch to its neighbors' sketches.}}
    \EndWhile
    \While { $|S|<K$ }
        \State BARRIER 
        \State $sums \leftarrow ${\sc Sketchwise-Sum}$(M)$\label{ln:3}\Comment{\parbox[t]{.37\linewidth}{Estimate cardinality from sketches.}}
        \State $sums \leftarrow $REDUCE$(sums, +)$ TO device $\tau=0$
        \If{$\tau=0$}
        \State $s \leftarrow argmax(sums)$\Comment{\parbox[t]{.37\linewidth}{Select a seed vertex.}} 
        \EndIf
        \State BROADCAST $s$ FROM device $\tau=0$ TO ALL \label{ln:broadcast} 
        \State $S \leftarrow S \cup \{s\}$\label{ln:4}\Comment{\parbox[t]{.37\linewidth}{Commit vertex s to the seed set.}}
        \State {$Q \leftarrow \{s\} $,~$\hat{Q} \leftarrow \varnothing $}\label{ln:5}
            \While{$Q \neq \varnothing $}
                \State $blocks \leftarrow min(size(Q),B)$
                \State {{\sc Cascade{\tt <<<}$blocks,{\cal J}${\tt >>>}}(${G'}_{\tau},M,{\cal J}, Q, \hat{Q} $)}\Comment{\parbox[t]{.37\linewidth}{Determine the reachability.}}
                \State {$Q \leftarrow \hat{Q}$,~$\hat{Q} \leftarrow \varnothing$}\label{ln:6}
            \EndWhile
            \State{$localscore \leftarrow ${\sc Count}$(M,-1)$}\Comment{\parbox[t]{.37\linewidth}{Calculate influence scores.}}\label{ln:7}
  
        \State $score \leftarrow ${ALLREDUCE}$(localscore, +) / {(\mu\times \cal{J}} )$\label{ln:8}
        \If {$(score-oldscore))/score > e$}
            \State {$M \leftarrow ${\sc{Fill-Sketches{\tt <<<}$B,{\cal J}$}{\tt >>>}}$(M, {\cal J}, \tau)$}\Comment{\parbox[t]{.37\linewidth}{Reset sketches.}}
            \While {$M$ not converged}
            \State $M \leftarrow ${\sc Simulate{\tt <<<}$B,{\cal J}${\tt >>>}}$({G'}_{\tau},M,{\cal J})$\Comment{\parbox[t]{.37\linewidth}{Rebuild sketches.}}
            \EndWhile
                    \State $oldscore \leftarrow score$

        \EndIf
    \EndWhile
    \State \Return $S$
\end{algorithmic}
\end{algorithm}

\begin{figure}[!ht] 
    \centering
        \includegraphics[width=0.75\linewidth]{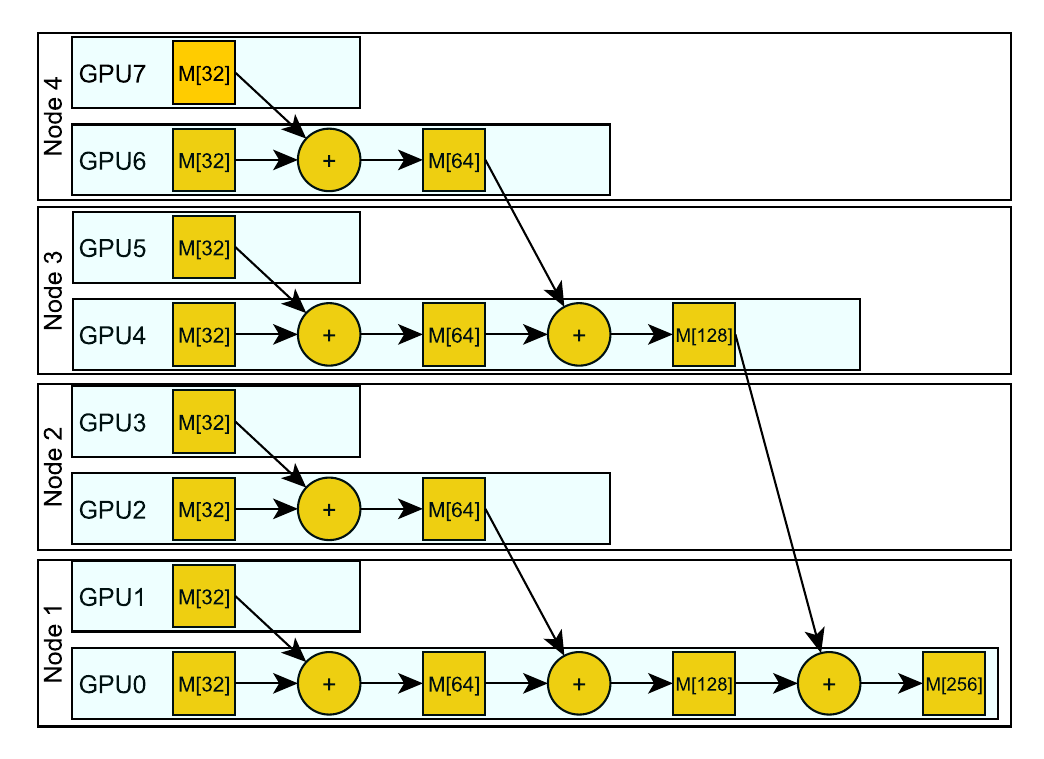}
  \caption{The reduction scheme of \acrod to find a seed vertex using sketch registers. The scheme uses 4 machines with 8 GPUs~(2 GPUs each), and ${\cal J}=256$ registers. M[$j$] represents a sketch with $j$ registers. The symbol $+$ represents the element-wise reduction of registers. 
  }
\label{fig:gpu-reduce}
\end{figure}

\acrod, whose pseudocode is given in Algorithm~\ref{algo:superfuser}, is built by using all the aforementioned kernels and techniques and consists of several distributed steps to efficiently estimate the reachability set cardinalities and select the seed vertices. First, all the devices start from their own random state which is an equisized partition of ${\cal X}$. For each partition ${\tau}$, where $0 \leq \tau < \mu$, we take only the sampled edges $E_{\tau}$; if the edge is sampled with any of the random values in the partition. We only copy these edges to the device-local graph ${G'}_{\tau} = (V, E_{\tau})$ of the corresponding task. The simulations are performed on this graph. Then, each device creates and initializes its own sketches for all vertices using its own random ${\cal X}$ values~(Algorithm~\ref{algo:fill-sketches}). Then it runs simulations on the sketches using the edges assigned to the partition/device~(with iterations given in Algorithm~\ref{algo:simulate-sketches}). 

\begin{figure}[!htbp] 
    \centering
    \includegraphics[width=0.75\linewidth]{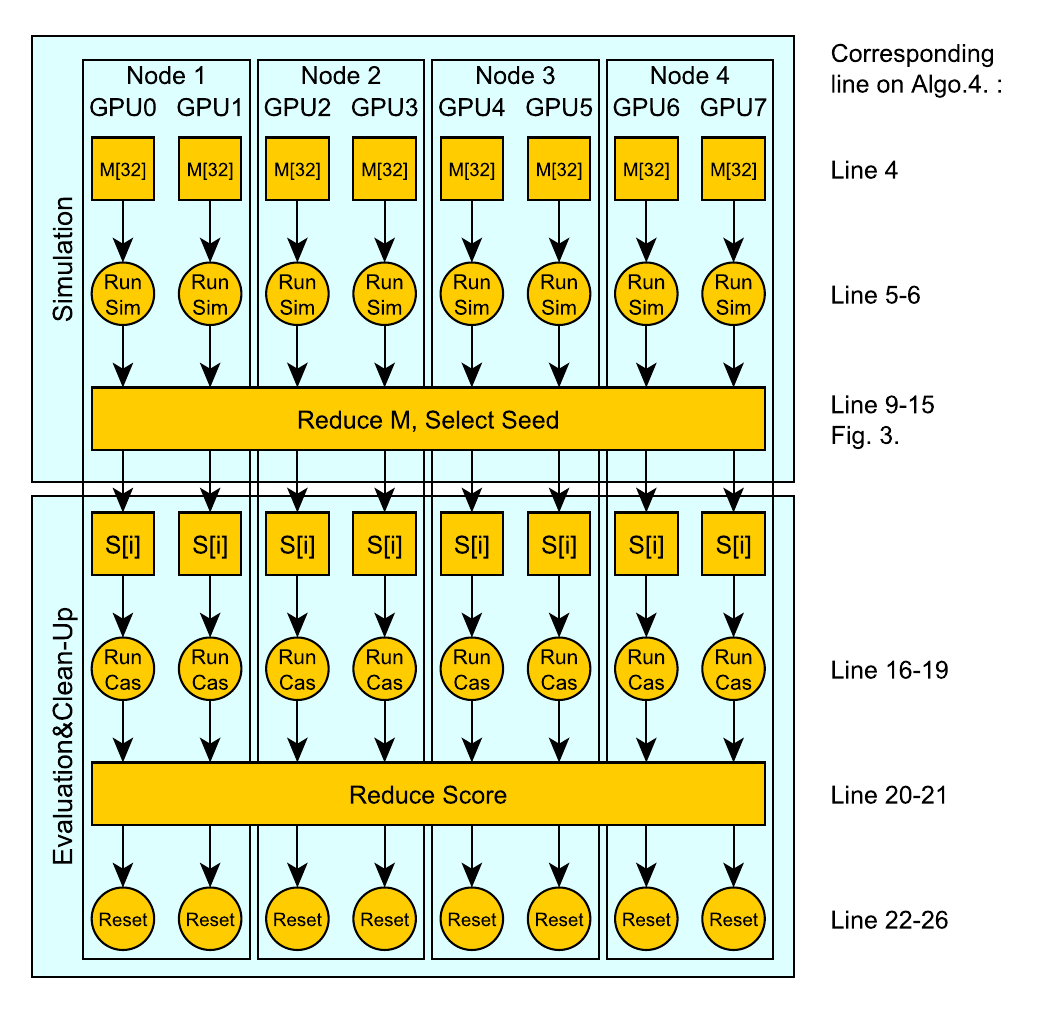}
    \caption{\label{fig:gpu-run} An example run of a seed selection in \acrod; The example scheme uses 4 machines with 8 GPUs~(2 GPUs each) and ${\cal J} = 256$ registers. M[$j$] represents a sketch with $j$ registers. Corresponding lines on Algo.4. are given on the right of the related steps. }
\end{figure}

After applying these preprocessing steps~(lines~\ref{ln:1}--\ref{ln:2} of Algorithm~\ref{algo:superfuser}), each device takes the harmonic mean of the register values to find the cardinality estimates of all candidate seed vertices. Then, an MPI reduction with $\log_2(\mu)$ levels is performed to aggregate the cardinality estimates on the leading device~($\tau=0$) where $\mu$ is the number of devices~(Fig.~\ref{fig:gpu-reduce}). The leading device selects the vertex with the highest estimated cardinality as the seed vertex as shown in the top part of Fig.~\ref{fig:gpu-run}. This seed is then broadcasted to all devices and added to the seed set $S$~(lines~\ref{ln:3}--\ref{ln:4} of Algorithm~\ref{algo:superfuser}).

After the seed selection, each device performs an influence cascade to calculate the influence of the (updated) seed set and label the visited vertices as -1~(lines~\ref{ln:5}--\ref{ln:6} of Alg.~\ref{algo:superfuser}). Finally, each device counts the number of visited vertices at its $\cal J$ simulations, i.e., the number of registers with value -1. These $localscore$ variables are reduced on all devices and averaged by dividing their sum to $\mu \times {\cal J}$ which is the total number of simulations to find the influence score~(lines~\ref{ln:7}--\ref{ln:8} of Alg.~\ref{algo:superfuser}).

For efficiency purposes, \acrod avoids expensive sketch updates, i.e., {\sc Simulate} iterations, if they do not have a significant impact on the overall influence. We observed that for many graphs, the first few seed vertices, where the number changes with the graph, have a significant impact on the influence. Since \acrod already estimates the marginal gains of the vertices, when they become insignificant, it uses the old sketch content to continue. That is when $$(score - oldscore))/score > e$$ for a given threshold $e$, \acrod does not perform simulations on the sketches. Here, $score$ is the influence of the current seed set whereas $oldscore$ is the influence score obtained after the last sketch rebuilt phase. We use $e = 0.01$ to avoid unnecessary updates while keeping the quality intact.

 Although Algorithm \ref{algo:superfuser} is parallel in most of the execution, it needs to explicitly synchronize all devices in a few places. 

 For instance, after running a random influence cascade, a barrier is required to estimate the scores correctly. Furthermore, $\log_2(\mu)+1$ BARRIERs are required for seed selection. 

\subsection{Fusing-aware Sample-Space Tasking}
\label{sec:fasst}
The hash-based generation of random values opens many performance opportunities. Since these values are generated using a random seed vector ${\cal X} = \{X_1, X_2,\ldots, X_{\cal R}\}$ for ${\cal R}$ simulations, we can manipulate, i.e., permute, the vector entries for better performance without losing randomness. Putting similar $X_r$ together reduces the path divergence, and hence, the number of ``wasted'' resources in SIMD lanes that do not contribute to the final result. It also reduces the (edge) overlap between two device-local graphs ${G'}_{\tau}$ and ${G'}_{\tau'}$.

In \acrod, a batch of samples for an edge is processed regardless of whether some samples do not include the edge. Hence, the threads in a warp diverge on consecutive samples; some threads perform the updates and the rest wait for joining to the next batch. When there is no divergence, all the warp threads will either perform an update or skip to the next batch.
As a simple yet effective approach to reduce the divergence, we opted to first sort ${\cal X}$ to reassign the simulation indices. Hence, similar bit-flips are clustered with respect to their significance. At one end of the sorted vector having small $X$ values, we expect the lower bits of the edge hash to be flipped. So the values for the same edge and for consecutive samples will be close to each other. At the other end of ${\cal X}$, we expect the higher bits to be flipped. For consecutive locations, the values are again expected to be close. This approach improves the fill rate of our SIMD lanes (on GPU, this corresponds to a warp) and allows us to provide an early exit for the edges which are selected in any of the samples in the same batch. 
Following this idea, we propose {\em Fusing-Aware Sample-Space Tasking}, FASST, while leveraging multiple GPUs. 
While distributing the samples into multiple tasks, the proposed method sorts ${\cal X}$ and takes advantage of the pseudo-random number generator. 
\begin{figure}[!htbp] 
    \centering
    \includegraphics[width=0.55\linewidth]{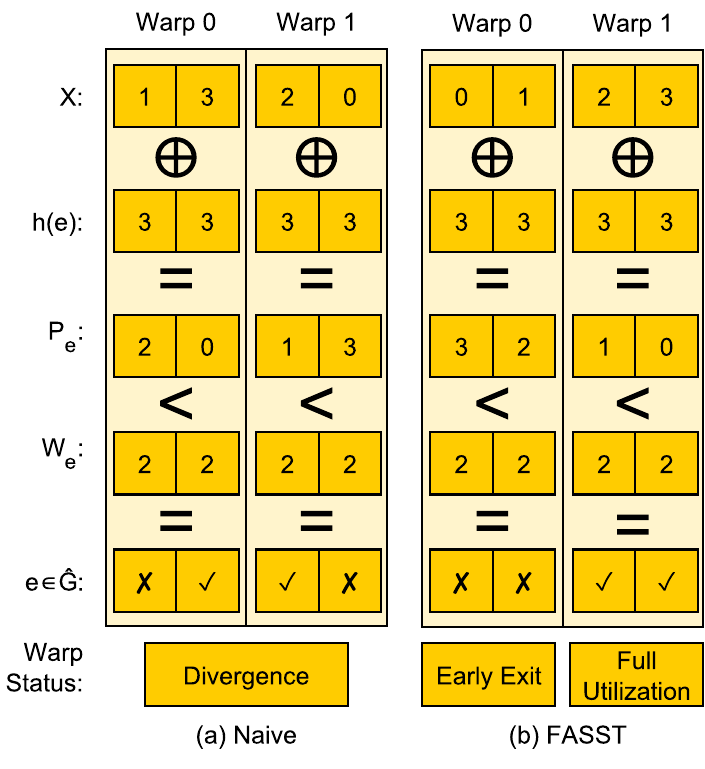}
    \caption{\label{fig:warp-divergence} An example run of fused sampling on edge $e$, using 4 samples and 2 threads wide (toy) warps. 
    Whether the edge $e$ is sampled, e.g.~$e \in \hat{G}$, is computed using the formula; $X \oplus h(e) < W_e$. $P_e$ and $W_e$ use a fixed-point representation. (a) uses naive random values where samples are evenly distributed amongst warps, causing divergent paths in warp 0 and 1. (b) uses FASST where either no edges are sampled or warp is fully utilized without any divergence. Warp 0 doesn't have any samples so an early exit is available. In Warp 1, all samples are active, which allows full utilization of the warp.}
\end{figure}

For small edge probabilities, only a small fraction of the edges can be active in a limited number of samples. Hence, the total number of edges to be processed can be extremely small compared to $|E|$. FASST allows us to efficiently cluster sampled and not sampled edges to simulation batches so that each task has fewer edges compared to the unsorted case, especially when the edge probabilities are small.

\section{Experimental Results} \label{sec:exps}
 The experiments are performed on four servers with two 64-core {\tt AMD EPYC 7742} processors, 1TB memory, and eight {\tt Nvidia A100}s. In total, on each node, there exist 128 cores for CPU computing, and each A100 has $6912$ {\tt CUDA cores} and $80$GB RAM. Due to the existing workload on the clusters, we were able to use up to $8\times$ {\tt Nvidia A100}s on 4 nodes, configured as 2 GPUs on each node.
 On the software side, {\tt CentOS 9.2} with {\tt GCC 9.2}, OpenMPI 3.0.2, and {\tt CUDA 11} are used to compile and run the experimented software. All the software is compiled using authors' build instructions. Our software uses optimization flags including; ``{\tt -use-fast-math}'', ``{\tt -march=native}'', and code generation is set for {Nvidia Ampere} architecture.

Experiments are performed on seven large social networks whose properties are given in Table~\ref{tab:NetPropsx}. All these networks are taken from SNAP network database~\cite{snapnets}. For all experiments, we set the seed set size as $K=50$ as in~\cite{MixGreedy}. For a comprehensive evaluation, we use five influence settings:
\begin{enumerate}
    \item constant edge weights $w = 0.005$; 
    \item constant edge weights $w = 0.01$~(as in~Kempe~et~al.\cite{kempe2003maximizing} and~Chen~et~al.\cite{MixGreedy});
    \item constant edge weights $w = 0.1$~(as in~Kempe~et~al.\cite{kempe2003maximizing});
    \item normally distributed edge weights with mean of $0.05$ and standard deviation of $0.025$, $w \in N(0.05,0.025)$;

    \item uniform distributed edge weights between $0$ and $0.1$, $w \in U(0,0.1)$

\end{enumerate}

\noindent We selected $w=0.005$ as a setting to challenge \acrod on extremely sparse samples. The later two settings,  $w = 0.01$  and $w = 0.1$, are selected to emulate the experiments of~\cite{MixGreedy} and~\cite{kempe2003maximizing}. In addition to constant edge weights, we performed two experiments with different statistical distributions to test how \acrod performs on mixed (large and small edge weights together) edge weights. For brevity, we represent constant sampling probabilities with their probability~(i.e., $0.005$, $0.01$, $0.1$ instead of $w=0.005$, $w=0.01$, $w=0.1$). Normal distribution samples $w \in N(0.05,0.025)$ are shown as $N0.05$ and Uniform distribution $w \in U(0,0.1)$ is shown as $U0.1$.

\begin{table}[tbp]
\centering
\caption{Properties of the graphs used in the experiments}\label{tab:NetPropsx}
\begin{tabular}{ll||r|r|r||l}
& & No. of           & No. of    & Avg.  &\\
&Dataset & Vertices          & Edges     &  Degree       & Type        \\
\hline
&{\tt LiveJournal} & 4,847,571 & 68,993,773  & 14.23 & Directed \\
&{\tt Pokec} & 1,632,803 & 30,622,564 & 18.75 & Directed\\
&{\tt Sinaweibo} & 58,655,849 & 261,321,071 & 4.46 & Directed\\
&{\tt Twitter} & 61,578,415 & 1,468,365,182  & 23.84  & Directed\\
\hline
&{\tt Friendster} & 65,608,366 & 1,806,067,135 & 27.53 & Undirected\\
&{\tt Orkut} & 3,072,441 & 117,185,083  & 38.14 & Undirected\\
&{\tt Youtube} & 1,134,891 & 2,987,625  & 2.63 & Undirected

\end{tabular}
\end{table}

\subsection{Choosing Baselines for Evaluation}

We compared the performance of \acrod with {\sc gIM}~\cite{gim} and {\sc cuRipples}~\cite{curipples}. {\sc gIM} is selected for being the fastest single-GPU algorithm and for its ability to process large datasets on a memory-restricted GPU. It is a recent, parallel, GPU implementation of {\sc IMM} algorithm~\cite{tang2015influence}. In our preliminary experiments, it was faster than {\sc cuRipples}. Also, according to Shahrouz~et~al.~\cite{gim}, {\sc gIM} is on-average $21.15\times$ faster than {\sc cuRipples} for seed set size $K=50$ with constant edge weights $w=0.05$. However,  {\sc gIM} cannot leverage multiple GPUs. On the other hand, {\sc cuRipples} is a scalable and distributed solution as shown in~\cite{curipples}.

We evaluate the algorithms on their execution times and influence scores since for IM, an algorithm must be fast and provide high-quality results to have merit. For instance, \acrod can compute relatively low-quality seed sets on the largest datasets without sketch rebuilding and be an order of magnitude faster. However, we needed to employ the rebuilding strategies so that it obtains high-quality seed sets while still being fast. To fairly compute the influence scores of the seed sets generated by the algorithms, we implemented a separate oracle. The scores are given as an average of three runs with different random seeds.

\subsubsection{Missing Values}

Unfortunately, we were not able to get results for some of the experiments due to the restricted memory on GPUs. Even though we have used 8x A100s each with 80 GBs of RAM, some experiments have failed experiments due to the high number of reachability sets, the sets of influenced vertices which \acrod does not generate but only estimate their cardinalities. 

For {\sc gIM}, the largest graph in the original paper is {\tt Orkut} with 117M edges~\cite{gim}. In our experiments, we have three larger graphs where two of which have more than a billion edges. In addition to the graph size, the number of required reachability sets increases when the sampling thresholds of the edges are small, e.g., $w = 0.005$ and $w = 0.01$ in our cases. Hence, for these cases, using GPUs for IM becomes harder for {\sc gIM}. For {\sc cuRipples}, the largest graph the authors used is again {\tt Orkut}~\cite{curipples}. This being said, our hardware resources were not on the same scale {\sc cuRipples} is tested on: the authors use up to 384 GPUs, and we have 8x A100 GPUs that can provide only 640 GBs of RAM in total. Due to these reasons, especially for larger graphs, both of the competitors lack many experimental values on Tables \ref{tab:single-per} and \ref{tab:multi-perf}. 

We note that the missing results do not over-emphasize \acrod's performance; in fact, they are likely to be long-running experiments that will probably increase the speedup \acrod would obtain. Unlike the baselines, ignoring the memory required to store the graphs once, the memory requirement of \acrod scales linearly with the number of vertices being processed and the number of simulations performed. The predictability of memory requirements allowed us to run all the experiments except one: unfortunately, the graph {\tt Friendster}, together with \acrod's sketches do not fit on the memory of a single GPU. This is why Table~\ref{tab:single-per} does not have any results for this graph. Only for the failed experiments, we modified our competitors' parameters (in good faith) to allow them to finish. For instance, whenever {\sc gIM} failed, we fine-tuned its reverse reachability set limits to make it run on more experiments. 

\acrod performs Monte-Carlo simulations to build seed sets' reachability set and calculate its own estimated influence scores which are used in the rebuilding step. Even though the reliability of the hash-based, fused-sampling method is shown by the experiments of G\"{o}kturk~and~Kaya\cite{infuser}, the number of Monte-Carlo simulations performed are fairly few in our settings. Due to this reason, we have used an independent oracle that does not have any optimizations and uses a large number of samples employing standard RNGs to verify the validity of the results.

\subsection{Comparing \acrod with {\sc gIM}}
The experiments show that compared to {\sc gIM}, \acrod can achieve high-quality results via only 1024 samples/registers to perform better in terms of both speed and quality. We have used $\epsilon=0.5$ for {\sc gIM}, which is stated as a sweet spot for the algorithm. This value aligns with the suggestion in Minutoli et al.~\cite{minutoli2019fast}, and it has also been proven effective in~\cite{gim}. 

Table~\ref{tab:single-per} compares the performance of the algorithms. The rightmost two columns show the influence scores whereas columns 3 and 4 show the execution times in seconds. The bottom two rows present the average and maximum improvement of \acrod(${\cal J}=1024$) over {\sc gIM}. On a single GPU, \acrod is $3.15\times$ faster than {\sc gIM} where the maximum speedup is around $8\times$. In terms of seed-set quality, \acrod performs better than {\sc gIM} in almost all settings. On average, {\sc gIM} can find seed sets that have $98\%$ influence of those of \acrod seed sets. 

\begin{table}[!ht]
\centering
\setlength{\tabcolsep}{3pt}
\caption{\label{tab:single-per} 
Execution times (secs.) and influence scores of \acrod~($\mathcal{J}=1024$) and {\sc gIM}~($\epsilon=0.5$) with $K = 50$ seeds. The last two rows are normalized w.r.t. {\sc gIM}. The experiment uses a single A100 GPU. Missing values due to out-of-memory errors are shown as ``-''.}
{
\begin{tabular}{ll||rr|rr}
 &  & \multicolumn{2}{c|}{Time} & \multicolumn{2}{c}{Score} \\\hline
Dataset & Setting & \acrod & {\sc gIM} & \acrod & {\sc gIM} \\\hline
{\tt Friendster} & - & - & - & - & -\\
\hline\multirow[c]{5}{*}{{\tt LiveJournal}} & 0.005 & \bfseries 78.86 & - & \bfseries 6022 & - \\
 & 0.01 & \bfseries 13.26 & - & \bfseries 48555 & - \\
 & 0.1 & \bfseries 10.08 & 49.58 & \bfseries 1700220 & 1695947 \\
 & N0.05 & \bfseries 11.47 & 52.29 & \bfseries 931219 & 925726 \\
 & U0.1 & \bfseries 10.76 & 52.27 & \bfseries 926163 & 920881 \\
\hline\multirow[c]{5}{*}{{\tt Orkut}} & 0.005 & \bfseries 122.69 & - & \bfseries 23025 & - \\
 & 0.01 & 129.33 & \bfseries 113.68 & 119506 & \bfseries 120913 \\
 & 0.1 & \bfseries 17.97 & 77.85 & \bfseries 2257592 & 2251873 \\
 & N0.05 & \bfseries 29.84 & 83.74 & \bfseries 1493737 & 1472394 \\
 & U0.1 & \bfseries 34.03 & 83.90 & \bfseries 1487416 & 1466698 \\
\hline\multirow[c]{5}{*}{{\tt Pokec}} & 0.005 & \bfseries 33.61 & - & \bfseries 601 & - \\
 & 0.01 & \bfseries 33.92 & - & \bfseries 1788 & - \\
 & 0.1 & \bfseries 3.35 & 26.72 & \bfseries 775088 & 774365 \\
 & N0.05 & \bfseries 3.95 & 29.64 & \bfseries 464404 & 463059 \\
 & U0.1 & \bfseries 4.17 & 29.61 & \bfseries 462174 & 460907 \\
\hline\multirow[c]{5}{*}{{\tt Sinaweibo}} & 0.005 & \bfseries 451.02 & - & \bfseries 792 & - \\
 & 0.01 & \bfseries 450.66 & - & \bfseries 4150 & - \\
 & 0.1 & \bfseries 295.59 & - & \bfseries 160684 & - \\
 & N0.05 & \bfseries 448.50 & - & \bfseries 94149 & - \\
 & U0.1 & \bfseries 455.35 & - & \bfseries 93499 & - \\
\hline\multirow[c]{5}{*}{{\tt Twitter}} & 0.005 & \bfseries 1804.24 & - & \bfseries 1620424 & - \\
 & 0.01 & \bfseries 1730.11 & - & \bfseries 3357055 & - \\
 & 0.1 & \bfseries 587.25 & - & \bfseries 18363884 & - \\
 & N0.05 & \bfseries 201.12 & - & \bfseries 2410378 & - \\
 & U0.1 & \bfseries 804.27 & - & \bfseries 12796400 & - \\
\hline\multirow[c]{5}{*}{{\tt Youtube}} & 0.005 & \bfseries 6.44 & 13.25 & \bfseries 1299 & 1141 \\
 & 0.01 & \bfseries 6.71 & 10.16 & \bfseries 3175 & 2865 \\
 & 0.1 & \bfseries 3.15 & 7.75 & \bfseries 131365 & 130372 \\
 & N0.05 & \bfseries 5.30 & 9.94 & \bfseries 54384 & 54097 \\
 & U0.1 & \bfseries 5.81 & 10.05 & \bfseries 53934 & 53822 \\
\hline \multicolumn{2}{l||}{Geo. Mean Performance} &\bfseries $3.15\times$ & $1.00\times$ & $1.02\times$ & $1.00\times$ \\
\multicolumn{2}{l||}{Max. Performance} &\bfseries $7.98\times$ & $1.00\times$ & $1.14\times$ & $1.00\times$ \\
 \end{tabular}
}
\end{table}

\begin{table}[!ht]
\centering
\setlength{\tabcolsep}{1.5pt}
\caption{\label{tab:multi-perf} 
Execution times (secs.) and influence scores of \acrod~($\mathcal{J}=1024$) and {\sc cuRipples}~($\epsilon=0.5$) with $K = 50$ seeds. The last two rows are normalized w.r.t. {\sc cuRipples}. The experiment uses 8 GPUs on 4 nodes, 2 GPUs on each. Missing values due to out-of-memory errors are shown as ``-''.}
{
\begin{tabular}{ll||rr|rr}

{} & {} & \multicolumn{2}{c|}{Time} & \multicolumn{2}{c}{Score} \\\hline
{Dataset} & {Setting} & {\acrod} & {{\sc cuRipples}} & {\acrod} & {{\sc cuRipples}} \\
\hline\multirow[c]{5}{*}{{\tt Friendster}} & 0.005 & \bfseries 186.93 & - & \bfseries 23982 & - \\
 & 0.01 & \bfseries 168.72 & - & \bfseries 790773 & - \\
 & 0.1 & \bfseries 101.17 & - & \bfseries 32457456 & - \\
 & N0.05 & \bfseries 100.80 & - & \bfseries 22228858 & - \\
 & U0.1 & \bfseries 124.37 & - & \bfseries 22159666 & - \\
\hline\multirow[c]{5}{*}{{\tt LiveJournal}} & 0.005 & \bfseries 6.63 & 48.89 & 6022 & \bfseries 6051 \\
 & 0.01 & \bfseries 4.66 & 72.79 & \bfseries 48555 & 48478 \\
 & 0.1 & \bfseries 3.81 & 38.59 & \bfseries 1700220 & 1699910 \\
 & N0.05 & \bfseries 3.61 & 26.83 & \bfseries 931219 & 930702 \\
 & U0.1 & \bfseries 3.63 & 21.98 & \bfseries 926163 & 924236 \\
\hline\multirow[c]{5}{*}{{\tt Orkut}} & 0.005 & \bfseries 5.96 & 36.78 & 23025 & \bfseries 23819 \\
 & 0.01 & \bfseries 6.24 & 31.94 & 119506 & \bfseries 122045 \\
 & 0.1 & \bfseries 2.78 & - & \bfseries 2257592 & - \\
 & N0.05 & \bfseries 3.63 & - & \bfseries 1493737 & - \\
 & U0.1 & \bfseries 3.25 & 48.12 & 1487416 & \bfseries 1488047 \\
\hline\multirow[c]{5}{*}{{\tt Pokec}} & 0.005 & \bfseries 2.76 & 54.56 & \bfseries 601 & 598 \\
 & 0.01 & \bfseries 2.97 & 36.54 & 1788 & \bfseries 1884 \\
 & 0.1 & \bfseries 1.53 & 357.44 & \bfseries 775088 & 774985 \\
 & N0.05 & \bfseries 1.43 & 10.68 & \bfseries 464404 & 463791 \\
 & U0.1 & \bfseries 1.56 & 8.04 & \bfseries 462174 & 461666 \\
\hline\multirow[c]{5}{*}{{\tt Sinaweibo}} & 0.005 & \bfseries 69.55 & - & \bfseries 792 & - \\
 & 0.01 & \bfseries 68.91 & - & \bfseries 4150 & - \\
 & 0.1 & \bfseries 61.12 & - & \bfseries 160684 & - \\
 & N0.05 & \bfseries 71.95 & 1689.49 & 94149 & \bfseries 94458 \\
 & U0.1 & \bfseries 71.86 & 1615.50 & 93499 & \bfseries 93753 \\
\hline\multirow[c]{5}{*}{{\tt Twitter}} & 0.005 & \bfseries 153.92 & - & \bfseries 1620424 & - \\
 & 0.01 & \bfseries 128.64 & - & \bfseries 3357055 & - \\
 & 0.1 & \bfseries 100.74 & - & \bfseries 18363884 & - \\
 & N0.05 & \bfseries 62.42 & 114.29 & \bfseries 2410378 & 2408968 \\
 & U0.1 & \bfseries 75.77 & - & \bfseries 12796400 & - \\
\hline\multirow[c]{5}{*}{{\tt Youtube}} & 0.005 & \bfseries 1.79 & 44.63 & \bfseries 1299 & 1292 \\
 & 0.01 & \bfseries 1.81 & 37.13 & \bfseries 3175 & 3152 \\
 & 0.1 & \bfseries 1.40 & 20.58 & \bfseries 131365 & 130843 \\
 & N0.05 & \bfseries 1.71 & 12.40 & \bfseries 54384 & 54335 \\
 & U0.1 & \bfseries 1.77 & 28.09 & \bfseries 53934 & 53367 \\
\hline \multicolumn{2}{l||}{Geo. Mean Performance} &\bfseries $11.99\times$ & $1.00\times$ & $1.00\times$ & $1.00\times$ \\
\multicolumn{2}{l||}{Max. Performance} &\bfseries $233.62\times$ & $1.00\times$ & $1.01\times$ & $1.00\times$ \\
 \end{tabular}
}
\end{table}

\subsection{Comparing \acrod with {\sc cuRipples}}
In the single-GPU experiments, we observed that \acrod can obtain high-quality results with ${\cal J}=1024$ registers in terms of solution quality. Hence, we keep the same setting for the multi-GPU experiments whereas we use $\epsilon=0.5$ for {\sc cuRipples} as also suggested in Minutoli~et~al.~\cite{minutoli2019fast}. As Table~\ref{tab:multi-perf} shows, overall, distributed \acrod, is on average $12\times$ faster than {\sc cuRipples} whereas the maximum speedup achieved is $233\times$. Since the speedups vary greatly, the geometric mean of performance metrics is used instead of the larger arithmetic mean. Furthermore, in terms of the seed set quality, Table~\ref{tab:multi-perf} shows that \acrod is as good as {\sc cuRipples} in almost all graphs and sampling probability settings.

\subsection{Impact of Sorting ${\cal X}$}

Sorting the random values in $\cal{X}$ results in higher utilization of threads and reduces thread divergence in \acrod. Setting ${\cal J} = 1024$ and using one thread per register result in 128 concurrent warps active on the same edge since a warp contains 32 threads. When there is branching, the threads which do not take the current path stay idle which can drastically reduce the performance. When $\cal{X}$ is sorted, similar sampling probabilities and hence bit flips are clustered together, so the processed edge is likely to be sampled by the consecutive threads.

As mentioned before, in the multi-GPU setting, \acrod generates a local graph for each device. By using a sorted $\mathcal{X}$, the Fusing-aware Sample Space Tasking (FASST) method drastically decreases the number of edges duplicated across these graphs. Furthermore, since this increases the chance of an edge being sampled on consecutive samples, each local graph will be more GPU-friendly; when a thread uses one of its edges, the probability of all the warp threads use the same edge is high. In fact, our experiments with low probability thresholds show that most of the edges are copied to only one graph. In addition, on dense samples~($w=0.1$), two copies are generated for each edge instead of eight (with 8 GPUs). The effect of FASST and sorting $\mathcal{X}$ on edge duplication can be found in Table \ref{tab:duplicates}.

Clustering similar samples together allows \acrod to sustain high GPU utilization and hide redundant, yet idempotent, operations that \acrod requires to overcome synchronization overhead. Considering only GPU kernels that contribute to results and excluding the initialization, \acrod has up to 86\% GPU utilization, whereas {\sc GIM} and {\sc curipples} only use 72\% and 50\% utilization respectively in our experiments on Orkut($w=0.01$) dataset. Even though high utilization is a good metric on dense computations for performance, on IM problem where computations are extremely sparse, utilization statistics can be misleading; \acrod implementation assumes graph traversal and updates are the bottlenecks and performs computations on hashing and fused sampling liberally. Arguably, \acrod is efficient in terms of occupancy, but wasteful in terms of resource consumption~(i.e., power draw).

Table \ref{tab:vector-fillrate} shows the average warp fill rate, which is the percentage of useful work throughout the execution, for both the naive method and FASST. The fill rate is calculated as follows; for all warps, we count the total number of threads in a warp that samples and uses an edge. Then, we divide this number by the total number of warp threads in which at least one edge is sampled. As the table shows, even with small sampling thresholds, i.e., $w = 0.005$, we double the useful work ratio, from 3\% to 6\%. For denser samples, e.g., when the edge weights are set to $w = 0.1$, we nearly quadruple the useful work done.
\begin{table}[tbp]
\setlength{\tabcolsep}{2pt}
\caption{\label{tab:duplicates} The avg. number of appearances for the  edges in $G$ in the device-local graphs. The averages are calculated across all the datasets where $\mathcal{J} = 1024$ samples are used.}

    \centering
{
\begin{tabular}{l|l||rrrrrrrrr}
      &   &  \multicolumn{9}{c}{\# of appearances}\\\hline
    Setting & Method &   0 &   1 &   2 &   3 &   4 &   5 &   6 &   7 &   8    \\
\midrule
           \multirow{2}{*}{$0.005$} &  Naive & 28\% & 37\% & 26\% &  8\% &  1\% &  - &  - &  - &  - \\
            &  FASST & 28\% & 71\% &  2\% &  - &  - &  - &  - &  - &  - \\
\midrule
            \multirow{2}{*}{$0.01$} &  Naive &  7\% & 23\% & 32\% & 26\% & 10\% &  3\% &  - &  - &  - \\
             &  FASST &  7\% & 86\% &  6\% &  - &  - &  - &  - &  - &  - \\
\midrule
             \multirow{2}{*}{$0.1$} &  Naive &  - &  - &  - &  - &  - &  - & 12\% & 23\% & 64\% \\
              &  FASST &  - & 17\% & 83\% &  - &  - &  - &  - &  - &  - \\
\midrule
\multirow{2}{*}{$N0.05$} &  Naive &  3\% &  1\% &  2\% &  5\% & 10\% & 13\% & 18\% & 22\% & 25\% \\
 &  FASST &  3\% & 74\% & 23\% &  - &  - &  - &  - &  - &  - \\
\midrule

     \multirow{2}{*}{$U0.1$} &  Naive &  4\% &  5\% &  5\% &  6\% &  7\% &  9\% & 16\% & 19\% & 30\% \\
      &  FASST &  4\% & 61\% & 34\% &  2\% &  - &  - &  - &  - &  - \\
\end{tabular}
}

\end{table}

\begin{table}[tbp]
\caption{\label{tab:vector-fillrate} Average warp fill rates~(percentage of threads that are contributing to the result) for each edge access with and without FASST.}

    \centering

\begin{tabular}{l||rr}
Setting & Naive &  FASST \\
\midrule
$w=0.005$           &    3\% &    6\% \\
$w=0.01$            &    4\% &    8\% \\
$w=0.1$             &   10\% &   44\% \\
$w\in N(0.05,0.025)$&    7\% &   31\% \\
$w\in U(0,0.1)$     &    7\% &   29\% \\
\end{tabular}
\end{table}

Table~\ref{tab:super-splits} shows the effect of splitting 1024 samples into two, four, and eight batches with naive distribution and FASST. As mentioned in Section~\ref{sec:fasst}, skipping edges that are known to be not sampled, can dramatically reduce the total number of edges on device-local graphs. Since the total time spent depends on the longest-running worker, we report the largest percentage of edges on a device-local graph in the table. The results show that for small edge probabilities such as $w \in \{0.005, 0.01\}$, even naive pre-sampling and skipping reduce the number of edges processed by each worker. Yet, FASST still manages to reduce the largest graph size by $4\%-12\%$ on average in these settings. For larger sampling thresholds, e.g., $w = 0.1$, we observe up to $3\times$ reduction on the device-local edge count when the graph is partitioned to eight. 

\begin{table}[tbp]
\setlength{\tabcolsep}{1.5pt}

\caption{\label{tab:super-splits} The largest percentage of edges on a device-local graph with and without FASST for 2, 4, and 8 devices. The averages are obtained over all the graphs.}

    \centering
\begin{tabular}{l||rr|rr|rr}
& \multicolumn{6}{c}{Number of devices} \\
 & \multicolumn{2}{c|}{2} & \multicolumn{2}{c|}{4} & \multicolumn{2}{c}{8}  \\\hline
Setting  & Naive & FASST & Naive & FASST & Naive & FASST \\
\midrule
$w=0.005$            &   48\% &   36\% &   29\% &   20\% &   15\% &   11\% \\
$w=0.01$             &   73\% &   46\% &   49\% &   26\% &   29\% &   15\% \\
$w=0.1$              &  100\% &   60\% &  100\% &   38\% &  100\% &   34\% \\
$w\in N(0.05,0.025)$ &   95\% &   55\% &   91\% &   31\% &   82\% &   19\% \\
$w\in U(0,0.1)$      &   92\% &   53\% &   86\% &   34\% &   76\% &   23\% \\

\end{tabular}
\end{table}

\subsection{Speedup with multiple GPUs}

Table \ref{tab:scaling-perf} shows how \acrod scales to multiple devices. The proposed method needs to assign at least ${\cal J} = 32$ samples for each GPU to utilize the warps efficiently. 
This being said \acrod may scale to more GPUs with ${\cal J} < 32$ if the graph and the sketches do not fit the GPU memory with ${\cal J} = 32$. In this case, multiple edges can be assigned to each warp which reduces warp-level communication but requires additional memory accesses in divergent paths. This limitation will be described in more detail later. 

Overall with 8 GPUs, \acrod achieves an average speedup of $7.09\times$, with a maximum speedup of $21\times$. Using the FASST method, speedups greater than the number of devices can be achieved by implicitly clustering similar samples which in turn uses much less memory as Table~\ref{tab:super-splits} shows.

This allows us to scale better on larger graphs than the state-of-the-art. Note that Table~\ref{tab:scaling-perf} does not have the graph {\tt Friendster} since the data structures of \acrod, graph and sketches, do not fit into a single GPU. 

 \begin{table}[!ht]
\centering
\setlength{\tabcolsep}{3pt}
\caption{\label{tab:scaling-perf} 
Speed-up achieved by \acrod using multiple GPUs ($K = 50$ and $\mathcal{J} = 1024$). 2 GPUs per node are used except for single GPU experiments. }

{
\begin{tabular}{ll||rrrr}

        & {} & \multicolumn{4}{c}{\# of GPUs (2 per node)} \\
      Dataset & Setting &       \multicolumn{1}{c}{1} &        \multicolumn{1}{c}{2} &        \multicolumn{1}{c}{4} &        \multicolumn{1}{c}{8}\\\hline
\multirow{5}{*}{{\tt LiveJournal}} & 0.005 &  $1.00\times$ &  $3.16\times$ &   $7.98\times$ &  $11.90\times$ \\
        & 0.01 &  $1.00\times$ &  $2.68\times$ &   $3.33\times$ &   $2.85\times$ \\
        & 0.1 &  $1.00\times$ &  $2.07\times$ &   $2.77\times$ &   $2.65\times$ \\
        & N0.05 &  $1.00\times$ &  $2.18\times$ &   $3.14\times$ &   $3.17\times$ \\
        & U0.1 &  $1.00\times$ &  $2.14\times$ &   $3.04\times$ &   $2.96\times$ \\
\cmidrule{1-6}
\multirow{5}{*}{{\tt Orkut}} & 0.005 &  $1.00\times$ &  $3.42\times$ &  $10.26\times$ &  $20.58\times$ \\
        & 0.01 &  $1.00\times$ &  $3.40\times$ &  $10.34\times$ &  $20.74\times$ \\
        & 0.1 &  $1.00\times$ &  $2.14\times$ &   $4.11\times$ &   $6.46\times$ \\
        & N0.05 &  $1.00\times$ &  $2.52\times$ &   $5.23\times$ &   $8.21\times$ \\
        & U0.1 &  $1.00\times$ &  $3.01\times$ &   $6.18\times$ &  $10.48\times$ \\
\cmidrule{1-6}
\multirow{5}{*}{{\tt Pokec}} & 0.005 &  $1.00\times$ &  $3.24\times$ &   $8.43\times$ &  $12.20\times$ \\
        & 0.01 &  $1.00\times$ &  $3.21\times$ &   $8.18\times$ &  $11.41\times$ \\
        & 0.1 &  $1.00\times$ &  $2.09\times$ &   $2.14\times$ &   $2.19\times$ \\
        & N0.05 &  $1.00\times$ &  $2.25\times$ &   $2.66\times$ &   $2.76\times$ \\
        & U0.1 &  $1.00\times$ &  $2.36\times$ &   $2.94\times$ &   $2.67\times$ \\
\cmidrule{1-6}
\multirow{5}{*}{{\tt Sinaweibo}} & 0.005 &  $1.00\times$ &  $2.63\times$ &   $4.71\times$ &   $6.49\times$ \\
        & 0.01 &  $1.00\times$ &  $2.60\times$ &   $4.97\times$ &   $6.54\times$ \\
        & 0.1 &  $1.00\times$ &  $2.39\times$ &   $3.53\times$ &   $4.84\times$ \\
        & N0.05 &  $1.00\times$ &  $2.49\times$ &   $4.52\times$ &   $6.23\times$ \\
        & U0.1 &  $1.00\times$ &  $2.52\times$ &   $4.60\times$ &   $6.34\times$ \\
\cmidrule{1-6}
\multirow{5}{*}{{\tt Twitter}} & 0.005 &  $1.00\times$ &  $2.97\times$ &   $7.32\times$ &  $11.72\times$ \\
        & 0.01 &  $1.00\times$ &  $2.90\times$ &   $6.84\times$ &  $13.45\times$ \\
        & 0.1 &  $1.00\times$ &  $1.88\times$ &   $4.75\times$ &   $5.83\times$ \\
        & N0.05 &  $1.00\times$ &  $1.68\times$ &   $2.65\times$ &   $3.22\times$ \\
        & U0.1 &  $1.00\times$ &  $2.92\times$ &   $5.64\times$ &  $10.61\times$ \\
\cmidrule{1-6}
\multirow{5}{*}{{\tt Youtube}} & 0.005 &  $1.00\times$ &  $2.16\times$ &   $3.04\times$ &   $3.60\times$ \\
        & 0.01 &  $1.00\times$ &  $2.19\times$ &   $3.17\times$ &   $3.71\times$ \\
        & 0.1 &  $1.00\times$ &  $1.80\times$ &   $2.00\times$ &   $2.25\times$ \\
        & N0.05 &  $1.00\times$ &  $1.83\times$ &   $2.49\times$ &   $3.10\times$ \\
        & U0.1 &  $1.00\times$ &  $2.00\times$ &   $2.65\times$ &   $3.27\times$ \\
\hline
\multicolumn{2}{l||}{Geo. Mean Performance} & $1.00\times$& $2.45\times$& $4.28\times$& $5.64\times$\\
\multicolumn{2}{l||}{Max. Performance} &  $1.00\times$& $3.42\times$& $10.34\times$& $20.74\times$\\
 \end{tabular}
}
\end{table}

Even though sketch building and diffusion steps are data independent between work units, \acrod needs to communicate cardinality estimations to select seed vertices for each sketch rebuild operation. The message preparation overhead is negligible since sketches are densely stored and all data is contigiously stored. 
Hence, the seed selection step performance is bound by communication throughput. To reduce the size of the communicated data, \acrod first performs a work unit local reduction that partially computes the cardinality for each vertex. So that, $\mathcal{O}(n)$ items are reduced instead of $\mathcal{O}(n\mathcal{J})$. Also, the implementation used {\tt CUDA-aware MPI} to circumvent extra overheads and allow memory access directly between GPUs over {\tt InfiniBand} without extra copies on main memory. Our experiments, in Table~\ref{tab:mpicost}, showed that only an extra 5.4\% of time spent at the worst case and most challenging and largest dataset in our experiments had the lowest communication overhead. Since the convergence of the diffusion kernel is slower for lower probabilities, e.g., 0.005, the percentage of the communication overhead is lower.

\begin{table}[!ht]
    \centering
    \caption{\label{tab:mpicost}Communication overheads w.r.t total run time in the same experiments on Table~\ref{tab:multi-perf}}
    \begin{tabular}{l||rrrrr}
         ~&  \multicolumn{5}{c}{Setting} \\ 
        \hline
        Dataset & 0.005 & 0.01 & 0.1 & N0.05 & U0.1 \\ 
        \hline
        
        {\tt Friendster}& 1.36\% & 1.51\% & 2.54\% & 2.55\% & 2.05\%  \\ 
        {\tt LiveJournal} & 2.87\% & 4.13\% & 5.10\% & 5.40\% & 5.37\%  \\ 
        {\tt Orkut} & 2.01\% & 1.91\% & 4.40\% & 3.34\% & 3.74\%  \\ 
        {\tt Pokec} & 2.31\% & 2.14\% & 4.24\% & 4.55\% & 4.16\%  \\ 
        {\tt Sinaweibo} & 3.32\% & 3.36\% & 3.80\% & 3.21\% & 3.21\%  \\ 
        {\tt Twitter} & 1.55\% & 1.86\% & 2.39\% & 3.91\% & 3.20\%  \\ 
        {\tt Youtube} & 2.48\% & 2.45\% & 3.19\% & 2.60\% & 2.51\%  \\ 
    \end{tabular}
\end{table}

\section {Possible Performance Limitations\label{sec:limitations}}
Even though \acrod can process the largest social graph we have in around 2-3 minutes and can generate high-quality seed sets, it requires multiple communication steps during its execution. At each step, an array of the same size as the number of vertices is communicated from each node/GPU. Since \acrod is extremely fast, we do not focus on this overhead in this study. However, when thousands of nodes are being used, this communication overhead can be alleviated by selecting important vertices and communicating only those vertices' scores or by employing a voting scheme to reduce data transfer between nodes.

As briefly mentioned above, another limitation of \acrod is how it distributes and schedules the samples. At its current state, it requires at least 32 samples/registers per device to work efficiently. 
This means \acrod is limited to work on only $\mathcal{J}/32$ devices. The problem can be partially solved by processing multiple vertices with a single warp and accepting small performance loss so that partitions smaller than ${\cal J} < 32$ can be processed. Using a small number of samples/registers per partition may decrease the effectiveness of FASST. For instance, in the extreme case, using a single register per partition increases the average number of memory accesses per sample, but does not decrease the number of edges per partition/device on some settings: with $w = 0.1$, FASST is able to reduce the size only $3.4\times$ with $32\times$ more GPUs. Hence, although an alternative implementation can use warp-level primitives to process multiple vertices/edges at once, the performance would not be comparable to \acrod. Due to this reason, we do not see a reason to add more hardware resources to the problem and accept the limitation as is. 

Another consideration is that we have focused on social networks where connections between vertices are distributed according to power law. \acrod implementation is optimized to perform well on social networks and assumes the reachability set forms a short-spanning tree. Networks with high vertex count with few edges and high diameters, such as road networks, are not considered in implementation. Performance on such networks can be improved by fusing multiple {\sc Cascade} kernel invocations to reduce the number of kernel launches. Also, implementing {\sc CELF}\cite{CELF} optimization to exploit monotone sub-modular selection criteria in seed selection can increase performance on networks with high vertex count. We have opted to skip these optimizations in this work to emphasize the proposed methods.  

\section{Related Work}\label{sec:rel}

There are other IM methods that can be considered cutting-edge or have  components similar to those of our work. Although there are studies using sketches, parallelism, or GPUs for IM in the literature, the methods proposed in this work differ in many ways. Sketch-based IM methods have less computational cost compared to simulation-based methods. Note that \acrod is a simulation-based method. In general, instead of repeatedly running simulations, sketch-based methods compute their sketches by pre-processing the influence graph to later assess the influence propagation. A popular method for sketch-based IM is {\sc Skim} by~Cohen~et~al.~\cite{cohen2014sketch} which uses combined sub-$k$ min-mixed accessibility plots built on $\ell$ sampled subplots to estimate the influence scores of the seed set~\cite{bottomk,cohen2015all}.  However, unlike \acrod and similar to most of the sketch-based approaches, {\sc Skim} does not match the quality of simulation-based methods. Furthermore, although the sketch utilization steps can run in parallel on a multicore CPU, the sketch creation step is single-threaded. 

Borgs~et~al.\cite{borgs2014maximizing} propose Reverse Influence Sampling~(RIS), which samples a fraction of all random reverse reachable sets. It then computes a set of $K$ seeds that covers the maximum number of sample vertices. The number of samples is calculated with respect to the number of visited vertices. The algorithm has an approximation guarantee of $(1-1/e-\epsilon)$. Minutoli~et~al.\cite{minutoli2019fast}~improved RIS and proposed {\sc IMM} that works on multi-threaded, distributed architectures and CUDA-capable GPUs. The CUDA accelerated implementation of the algorithm is branded as {\sc cuRipples}.

Shahrouz~et~al. proposed {\sc gIM}~(GPU-accelerated Influence Maximization) algorithm. {\sc gIM} is based on IMM~\cite{tang2015influence} method, and it improves the running time on large-scale graphs with low values of $\epsilon$. The reverse reachability generation is accelerated by work-efficient parallel processing of frontier vertices, and multiple breadth-first searches which are performed in parallel.

Also, even though it is not focusing on GPUs and the distributed setting, we have borrowed the hash-based fused-sampling from~\cite{infuser} and sketch-based estimation of the reachability sets from~\cite{hyperfuser}. However, we employ GPU-specialized sketching methods and scheduling optimization via ordering random values to increase performance.

Besides works on the IM problem, there are works on efficient partitioning for graph processing. HotGraph has been proposed by Zhang~et~al. By identifying a critical subgraph within the entire graph, HotGraph significantly reduces cross-partition communication overhead. Coupled with optimized partition scheduling and execution strategies, HotGraph demonstrates substantial improvements in both the number of vertex state updates and overall processing time.

\section{Conclusion}\label{sec:conc}
In this work, we propose a sketch-based IM algorithm that employs fusing and runs on multiple GPUs and nodes efficiently. We applied various techniques for an efficient implementation that utilizes fast instruction patterns in CUDA, maximizes GPU occupancy, and balances the use of memory units. The proposed algorithm concurrently processes multiple samples for better performance. In addition, we propose the Fusing-Aware Sample-Space Tasking method to partition the edges of the graph to multiple GPUs efficiently by sorting random seed values. Also, we present a performance comparison with state-of-the-art, GPU-based IM algorithms on real-world datasets and show that \acrod can be faster while providing high-quality results. The implementation is on average $12\times$ faster than the competition on 8 GPUs compared to a state-of-the-art, distributed IM tool.

\bibliography{refs}

\end{document}